# Phonon Scattering in the Complex Strain Field of a Dislocation

Yandong Sun[1,5], Yanguang Zhou[2], Ramya Gurunathan[5], Jin-Yu Zhang[1], Ming Hu[4], Wei Liu[1,*], Ben Xu[3,*], and G. Jeffrey Snyder[5,*]

[1]Laboratory of Advanced Materials, School of Materials Science and Engineering, Tsinghua University, Beijing 100084, People's Republic of China.

[2]Department of Mechanical and Aerospace Engineering, The Hong Kong University of Science and Technology, Hong Kong, People's Republic of China.

[3]Graduate School, China Academy of Engineering, Beijing 100193, People's Republic of China.

[4]Department of Mechanical Engineering, University of South Carolina, Columbia, SC29208, USA.

[5]Department of Materials Science and Engineering, Northwestern University, Evanston, IL 60208, USA.

*email: bxu@gscaep.ac.cn
weiliu@mail.tsinghua.edu.cn
jeff.snyder@northwestern.edu

## Abstract

Strain engineering is critical to the performance enhancement of electronic and thermoelectric devices because of its influence on the material thermal conductivity. However, current experiments cannot probe the detailed physics of the phonon-strain interaction due to the complex, inhomogeneous, and long-distance features of the strain field in real materials. Dislocations provide us with an excellent model to investigate these inhomogeneous strain fields. In this study, non-equilibrium molecular dynamics simulations were used to study the lattice thermal conductivity of PbTe under different strain status tuned by dislocation densities. The extended 1D McKelvey-Shockley flux method was used to analyze the frequency dependence of phonon scattering in the inhomogeneously strained regions of dislocations. A spatially resolved phonon dislocation scattering process was shown, where the unequal strain in different regions affected the magnitude and frequency-dependence of the scattering rate. Our study not only advances the knowledge of strain scattering of phonon propagation but offers fundamental guidance on optimizing thermal management by structure design.

## 1. Introduction

Over the past half-century, the pursuit of faster, cheaper computing and increased device density has driven the size of the processor down[1]. A minimum feature size of 2 nm is being targeted for next-generation processors[2]. High thermal conductivity for dissipating the heat generated by work processors in time is important to determine their speed, efficiency, and reliability[3]. On the other hand, in the thermoelectric field, lower thermal conductivity enables a higher energy conversion efficiency[4]. Thus, thermal management improves the performance of these electronic and thermoelectric devices. Atomic defects such as dislocations are introduced when the devices are produced, which will generate a complex strain field[5, 6]. Understanding the effect of the strain field on the lattice thermal conductivity is important to thermal management and also a very difficult research topic. Systematic experimental and computational studies of phonon scattering from strain fields have thus far been missing. One of the obstacles is the long-distance nature of the strain field, which is beyond the capability of the ab initio calculations due to their limited computational scale. Only uniaxial strain and hydrostatic pressure are studied,[7-10] which is a very simplified approximation. The second obstacle is that in real situations the strain has different components, which is illustrated in High-resolution TEM (HRTEM) observation[6, 11-13]. An inhomogeneous strain field is introduced for example by randomly distributed dislocations. However, only an average strain can be calculated using XRD diffraction[14], neutron scattering[15], and Raman scattering[16]. From the average strain, theoretical equations derived by Klemens [17, 18] and Carruthers[19] are used to predict the phonon-strain scattering rate. The effect of dislocation strain on the thermal properties of semiconductors has received considerable attention from researchers[6, 20-23]. The thermal properties of the defected model can be simulated by molecular dynamics (MD) simulation[24-31], which can capture the long distance nature of the strain field. The spectral heat flux determined through MD[27, 32] and the extended 1D McKelvey-Shockley flux method[33] can deconvolute the phonon scattering in different regions in thermal inhomogeneous materials. Thus it is possible to obtain a spatially resolved phonon dislocation scattering as a prototype for phonon scattering with lattice strain.

In this study, phonon scattering analysis is performed for the first time for both strain local to the dislocation core as well as non-local shear strain using non-equilibrium molecular dynamics (NEMD) simulations of a PbTe crystal with different dislocation densities. The $\kappa_L$ and spectral heat flux across a plane at the site of the dislocation, perpendicular to the Burgers vector, was directly obtained. The

$\kappa_L$ in dislocation models was significantly reduced compared to the result of a pristine structure model. The phonon scattering from the local region around the dislocation core and non-local shear strain region persisting throughout the model were separated using the extended 1D McKelvey-Shockley flux method in thermally inhomogeneous materials. This study not only demonstrates the variance in phonon scattering from the dislocation core and strain field but also provides an estimated phonon scattering rate in these two phonon scattering regions.

## 2. Models and Methodology

### 2.1 Models

A model of PbTe with a dislocation network was built using our in-house code as shown in Fig. 1(a). The model dimensions were 60, 60, and 20 nm in the X, Y, and Z directions, respectively. Randomly oriented dislocations were added to the crystal and the model was relaxed at 20 K with the isothermal-isobaric ensemble (NPT) in the MD simulations. The final structure was divided into three regions in the Z direction. The normal strain $\epsilon_{xx}$ and $\epsilon_{yy}$ and shear strain $\gamma_{xy}$ for a XY cross-section in each of the three regions from bottom to top were shown in Fig. 1(b), (c), and (d), respectively. Each region differs in terms of the intensities of the strain components as a result of the changes in dislocation distribution. The shear strain component was larger in magnitude as well as not only larger but also more widely resolved than normal strain in the crystal.

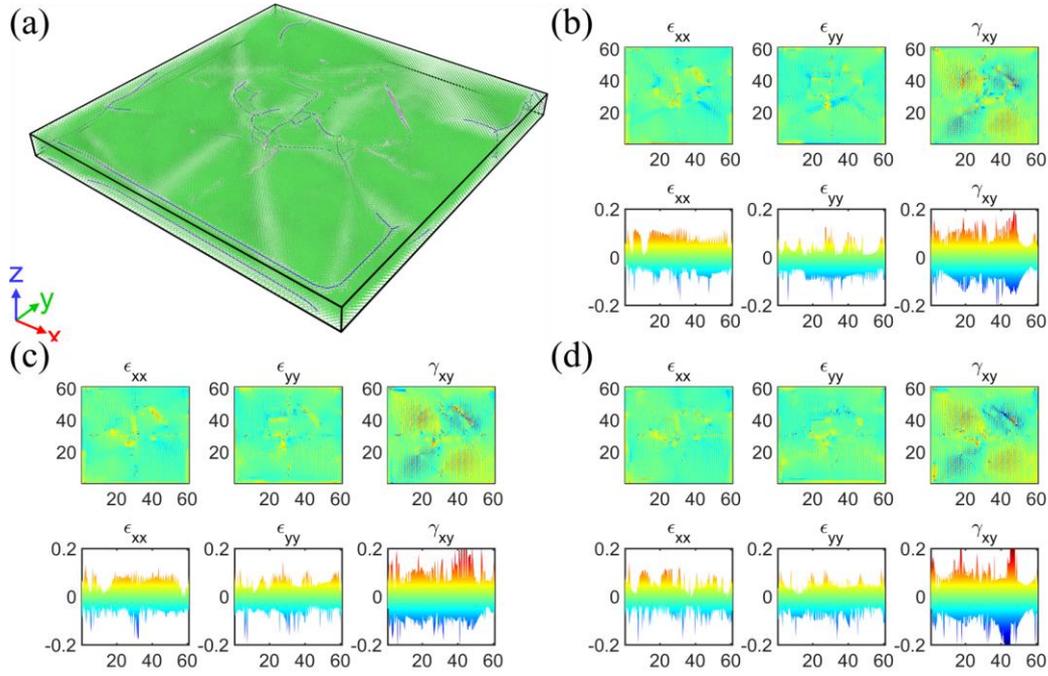

Fig. 1. The dislocation network model and its normal strain $\epsilon_{xx}$ and $\epsilon_{yy}$ and shear strain $\gamma_{xy}$. (a) Dislocation net model: blue lines are the dislocations mapped using Ovito software[34]. (b), (c), and (d) Strain fields of the three different XY plane layers in the Z direction.

To simplify the problem, a dislocation PbTe model (D Model) with only one isolated dislocation was built as shown in Fig. 2(a). The simulation contained 159,552 atoms arranged in a rock salt structure (lattice parameter $a$ = 6.447 Å at 20 K) oriented as $\bar{u}_x = [\bar{1}\ 1\ 0]$, $\bar{u}_y = [\bar{1}\ \bar{1}\ 2]$, and $\bar{u}_z = [1\ 1\ 1]$. The simulation dimensions were $63.15 \times 9.48 \times 8.94$ nm with an edge dislocation in the middle of the model with a Burgers vector of $\frac{a}{2}[\bar{1}\ 1\ 0]$, resulting in a dislocation density of $3.34 \times 10^{15}$ m$^{-2}$. Periodic boundary conditions were satisfied to eliminate boundary scattering, and dislocations were sufficiently distanced to prevent interactions. The D Model's strain field is plotted in Fig. 3. The normal strain $\epsilon_{xx}$ and $\epsilon_{yy}$, the two largest strain components, decreased from 10% to 0.1% in a range of 5.35 nm in the radius around the dislocation core (between the blue and red arrows in Fig. 3). In the other regions, the $\epsilon_{xx}$ and $\epsilon_{yy}$ were nearly zero (< 0.1%), except for a continuous $\gamma_{xy}$ (~2%). The same behavior is shown in the dislocation strain fields from continuum elasticity theory, where $\epsilon_{xx}$ is mainly distributed along the extra half plane and $\gamma_{xy}$ is mainly distributed along the Burgers vector direction[35]. The different distribution behavior of the individual strain components shows the origin of the two different scattering regions discussed previously, the LSR and NLSR. The LSR is produced by the elevated normal and shear strain approximately 5.0 nm around the dislocation core, while the NLSR is generated by the spatially extended shear strain. Different dislocation density models were also built. The dislocation density varied from $1.5 \times 10^{15}$ m$^{-2}$ to $6.5 \times 10^{15}$ m$^{-2}$ by changing the model's length in the Y direction.

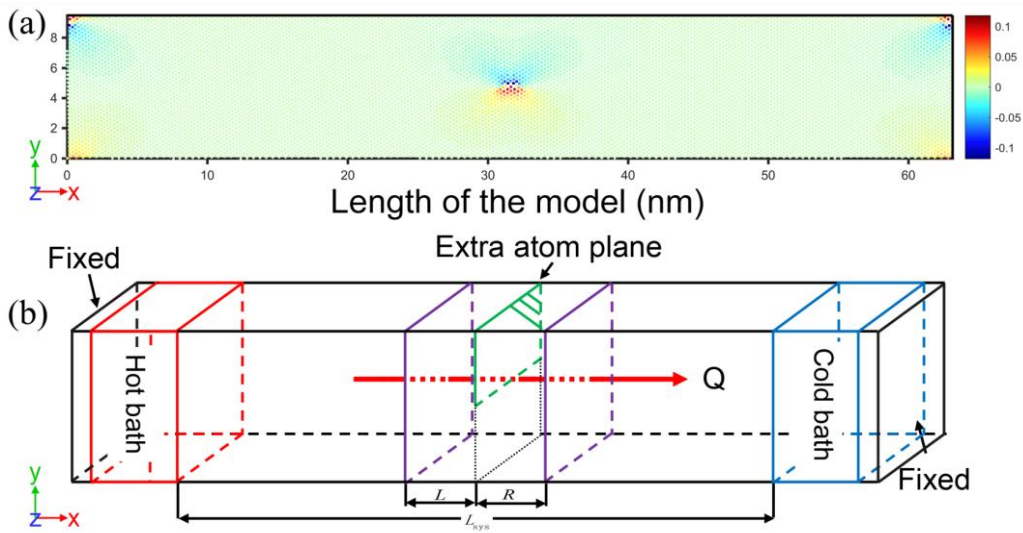

Fig. 2. (a) XY plane of the dislocation model with a $3.34 \times 10^{15}$ m$^{-2}$ dislocation density in the NEMD simulations colored by the normal strain field $\epsilon_{xx}$. The model was 63.15 nm long in the X direction and had a 9.48 nm × 8.94 nm cross-section. An edge dislocation was placed in the middle of the model with a Burgers vector of $\frac{a}{2}[\bar{1}\ 1\ 0]$. (b) Schematic of the simulation cell used in the

NEMD simulations.

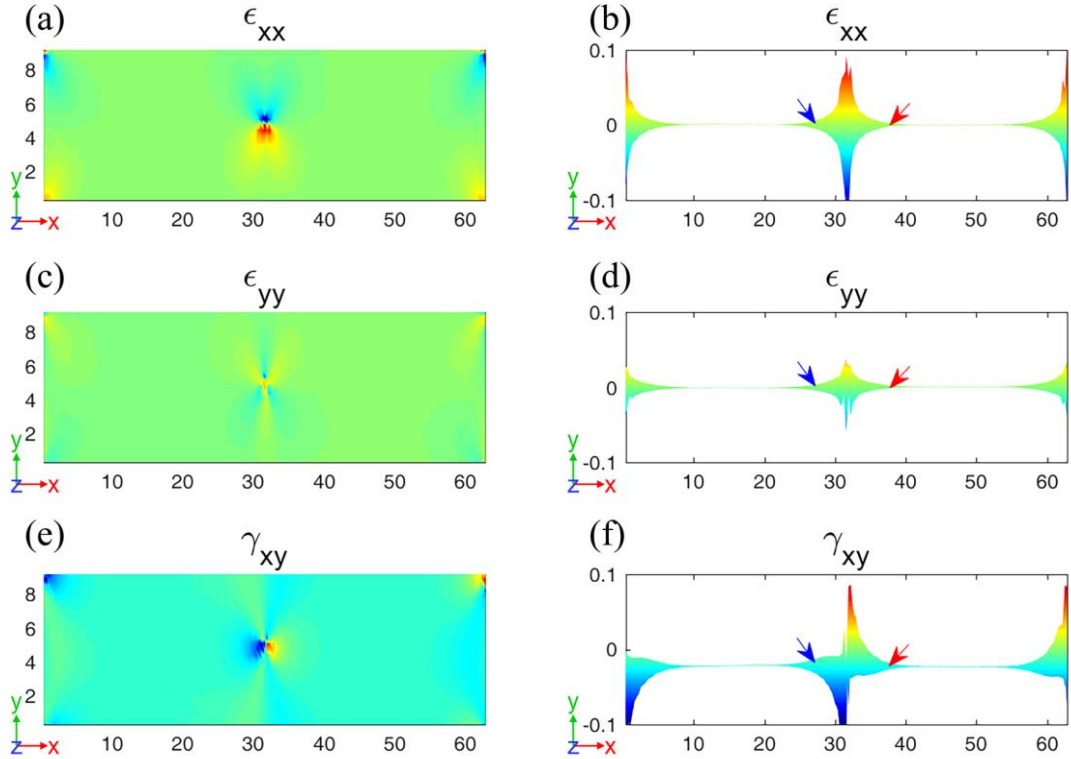

Fig. 3. The normal strain $\epsilon_{xx}$ and $\epsilon_{yy}$ and shear strain $\gamma_{xy}$ of the model with dislocation density $3.34 \times 10^{15}$ m$^{-2}$. (a) $\epsilon_{xx}$, (c) $\epsilon_{yy}$, and (e) $\gamma_{xy}$ show the strain variation on a cross-sectional XY plane coinicident with the dislocation core. (b) (d) and (f) show the 1D plots of each strain component along an X-direction scan at the center of the dislocation core. The $\epsilon_{xx}$, $\epsilon_{yy}$ at the dislocation core generate a local scattering region (LSR) and the extended $\gamma_{xy}$ generates a non-local scattering region (NLSR).

A schematic of the simulation cell used in the NEMD simulations is shown in Fig. 2(b). The atoms at each end of the system were fixed. Next to both the right and left fixed boundary, a 2 nm length bath contained atoms coupled to a Langevin thermostat with a damping parameter of 0.1 (20 times the time step) and the keywords "tally yes." The temperature was fixed at 40 K in the hot bath and 0.1 K in the cold bath, respectively. The dislocation's extra atomic plane (in green) was seen as a virtual interface, and the spectral heat flux through it was calculated . Two groups of atoms (denoted by L and R) were selected on both sides of the interface, and their velocities were sampled at successive time steps, which were used to calculate the spectral heat flux across the virtual interface. The same simulation procedure was used for the pristine structure model.

## 2.2 NEMD simulations

The LAMMPS[36] package was used to conduct the NEMD simulations with a time step of 5 fs. A Buckingham pair potential from Ref. [37] was used to model the interactions between atoms. A total of $4.5 \times 10^6$ NEMD steps were performed, corresponding to a running time of 28 ns. The first 12.5 ns was used to relax the structure with the isothermal-isobaric (NPT) and canonical (NVT) ensembles, then the following 15 ns was used to obtain a steady temperature gradient and heat flux. Langevin thermostats were applied to the hot and cold regions with a damping parameter of 0.1 (20 times the simulation's time step) and the keywords "tally yes" as shown in the schematic of the simulation cell in Fig. 2(b). The matrix between the hot and cold regions was coupled to the microcanonical (NVE) ensemble. The time-domain velocities of atoms for spectral heat flux calculations and temperature profiles were sampled for the last 1.5 ns.

## 2.3 Spectral heat flux method

The spectral heat flux $Q(\omega)$ was calculated using the following equation[27, 32]

$$Q(\omega) = -\frac{2}{t_{simu}\omega A}\sum_{i\in L, j\in R}\sum_{\alpha,\beta\in\{x,y,z\}}\text{Im}\langle \tilde{v}_i^\alpha(\omega)^* K_{ij}^{\alpha\beta} \tilde{v}_j^\beta(\omega)\rangle, \qquad (1)$$

where $t_{simu}$ is the total simulation time, $A$ is the area of the cross-section, $\alpha$ and $\beta$ are the Cartesian directions, $K_{ij}^{\alpha\beta}$ is the second-order interatomic force constant between the atoms resolved on both sides of the dislocation (denoted by $L$ and $R$), $\tilde{v}_i^\alpha(\omega)$ and $\tilde{v}_j^\beta(\omega)$ are the discrete Fourier transformed velocities of atoms $i$ in direction $\alpha$ and atoms $j$ in direction $\beta$, and $\tilde{v}_i^\alpha(\omega)^*$ is the complex conjugate of $\tilde{v}_i^\alpha(\omega)$. The bracket notation $\langle\rangle$ denotes the steady-state non-equilibrium ensemble average. The velocity of the atoms was directly obtained by the NEMD simulations and the interatomic force constants were obtained using the finite displacement method in the MD simulations.

## 2.4 Nominal phonon mean free paths

The phonon mean free paths (MFPs) $\lambda(\omega)$ can be obtained in thermally homogeneous crystals from spectral heat fluxes in NEMD simulations of different lengths, where length is the distance between the hot and cold baths[33]. In our previous work, this method, the 1D McKelvey-Shockley method, was extended to crystals with defects, which produce localized regions of increased scattering that result in thermal inhomogeneities observable in the simulation's temperature profile. The defects induce a local scattering region (LSR) with a MFP $\lambda_1(\omega)$ and a non-local scattering region (NSLR; scattering effects present throughout the model) with a MFP $\lambda_2(\omega)$[33]. As discussed in our previous work[33], only a nominal phonon MFP $\lambda_{nm}(\omega)$, which is a function of $\lambda_1(\omega)$ and $\lambda_2(\omega)$, can be

obtained from an inhomogeous model with defects. The expression for $\lambda_{nm}(\omega)$ is

$$\lambda_{nm}(\omega) = \lambda_2(\omega) + d\left(\frac{\lambda_2(\omega)}{\lambda_1(\omega)} - 1\right), \qquad (2)$$

where $d$ is the length of the LSR, which can be obtained from the temperature profile by determining the width of the anomalous temperature drop local to the defect (details in the Section 3.1). Of note, the $\lambda_{nm}(\omega)$ can not be interpreted as a real phonon MFP, but instead is a way to compare the inhomogeneous, defective crystal to the pristine case. The comparison between the phonon MFP of the pristine structure $\lambda_p(\omega)$ and $\lambda_{nm}(\omega)$ allows one to better understand the relationship between the LSR and NLSR scattering.

The phonon MFPs $\lambda_p(\omega)$ in the pristine structure model and nominal phonon MFPs $\lambda_{nm}(\omega)$ in the defect model are obtained by fitting length-dependent spectral heat fluxes[27, 32, 33]. Here, we will describe how comparing these values can provide a qualitative understanding of local and non-local scattering in the defect model. First, we will assume the phonon MFPs in the NLSR $\lambda_2(\omega)$ are equal to the phonon MFPs in the pristine structure model $\lambda_p(\omega)$. This assumption is not expected to be generally applicable, since dislocation strain scattering persists in the NLSR. However, it may apply to certain phonon frequencies. If a phonon with frequency $\omega'$ is scattered specifically in the LSR, then the phonon MFP in the LSR will be smaller than the phonon MFP in NSLR (i.e. $\lambda_1(\omega') < \lambda_2(\omega')$). As a result, $d(\lambda_2(\omega')/\lambda_1(\omega') - 1) > 0$ and $\lambda_{nm}(\omega') > \lambda_2(\omega') = \lambda_p(\omega')$. The phonon frequencies which experience increased scattering in the LSR are identified by noting the regions of the phonon spectrum where the nominal phonon MFPs $\lambda_{nm}(\omega)$ are greater than the pristine structure

MFP $\lambda_p(\omega)$. In the next case, we will relax the requirement of $\lambda_2(\omega) = \lambda_p(\omega)$, but instead assume that $\lambda_2(\omega)$ is roughly equal to $\lambda_1(\omega)$, indicating similar scattering in the LSR and NLSR. If a phonon with frequency $\omega''$ is scattered by non-local strain effects in the NLSR, then $\lambda_2(\omega'')$ will be smaller than $\lambda_p(\omega'')$, since the pristine model solely includes phonon-phonon scattering. In this case, where we assume similar scattering from the LSR and NSLR, $\lambda_{nm}(\omega'') = \lambda_2(\omega'') < \lambda_p(\omega'')$. Therefore, the phonon frequencies that are scattered by the NLSR can be identified as the part of the MFP spectrum where $\lambda_{nm}(\omega)$ is less than the pristine structure model $\lambda_p(\omega)$. However, if a phonon was scattered both by the LSR and NLSR together, the former scattering process will increase the $\lambda_{nm}(\omega)$, but the later one will decrease the $\lambda_{nm}(\omega)$, there is a possibility that the two changes are offset. So it is not possible to obtain an explicit frequency range in which the phonons are scattered by the LSR or NLSR.

## 3. Results and discussion

### 3.1 Temperature distribution and thermal conductivity

The temperature distribution along the heat flux direction in the dislocation D models and pristine structure P Model are shown in Fig. 4. The P Model had a linear temperature distribution. In contrast, the D models contained a temperature drop at the position of dislocation core, which increased in magnitude as the dislocation density increased. This drop occurs because of the heighted phonon resistance in the core region, which means a higher temperature gradient is required to maintain the same heat flux as in the other parts of the system. The temperatures drops were also seen near the baths, which indicates that some phonons were ballistically transported through the sample with no scattering.

As such, their phonon mean free paths were larger than the length of the models. The shape of the temperature profiles was can be explained by the McKelvey-Shockley phonon BTE method[38-40]. The ballistic phonons will not change the system's phonon population, thus generating a flat temperature distribution. Combined with the linear temperature distribution produced by diffusive phonons, the resulting profile contains temperature drops near the baths and a linear temperature distribution in the system. In the D Model, the temperature drops near the baths decreased as the dislocation density increased because some phonons underwent a transition from ballistic to diffusive transport as a result of the increased dislocation scattering.

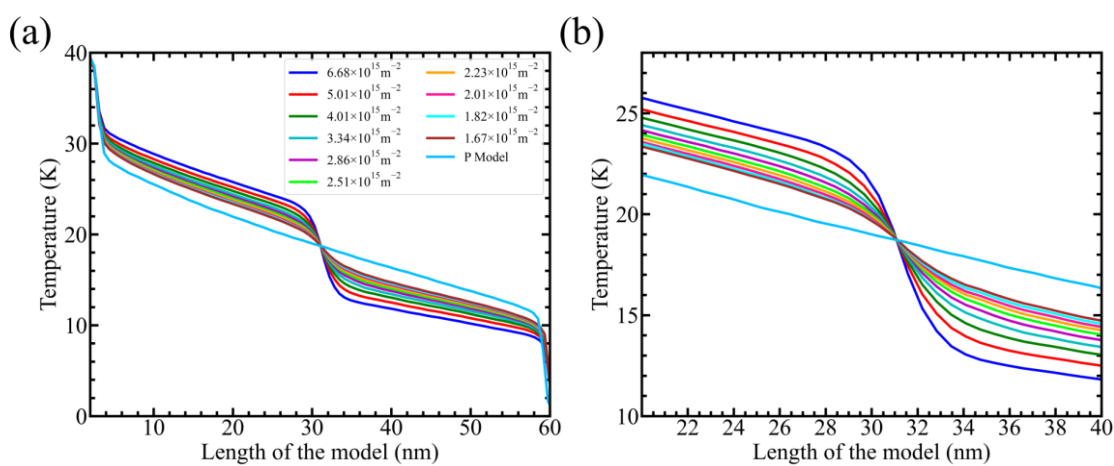

Fig. 4. (a) The temperature distributions of the dislocation models with various defect densities in addition to the pristine structure P Model in the NEMD simulations. (b) Magnified view showing of the sharp reduction in temperature near the dislocations.

The position where the temperature distribution begins to deviate from the linear distribution and the position where it returns to the linear distribution are indicated by blue and red arrows in the dislocation strain field as shown in Fig. 3. The phonons were severely scattered in the LSR. The length of the LSR can be obtained from the temperature distribution, see the example in Supplementary Fig. 1. However, it was not possible to resolve the scattering from the NLSR using the temperature distribution alone, because the temperature profile in the NLSR was significantly influenced by the LSR. Therefore, the 1D McKelvey-Shockley flux method, described in Section 2.4, was applied to qualitatively describe the scattering from the NLSR.

The lattice's thermal conductivity $\kappa_L$ was directly calculated using Fourier's law, and the results are shown in Fig. 5. The temperature gradient was calculated in the whole region. The $\kappa_L$ decreased hyperbolically as the dislocation density increased, which was also shown in Eqns. 20 and S59 of Ref.[41]. The $\kappa_L$ decreased by 56.9% in the highest dislocation density model compared to the P Model.

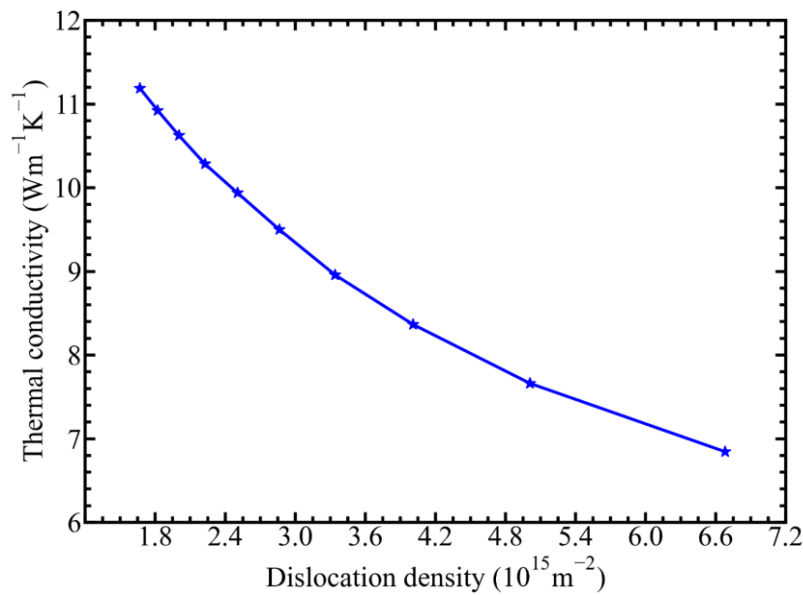

Fig. 5. The thermal conductivity decreased with the dislocation density. The thermal conductivity of pristine structure P Model was 15.9 W/(m·K).

## 3.2 Spectral heat flux

The spectral heat flux $Q(\omega)$ of the D Model and P Model, which is related to $\kappa_L$, was calculated as shown in Fig. 6(a). We get the total heat flux $Q_{\text{total}}$ by integrating $Q(\omega)$ over frequency. The contribution to the total heat flux from different frequency phonons was observed from the $Q(\omega)$. The contribution from phonons in a frequency range of 1.1 THz to 1.9 THz decreased as the dislocation density increased due to stronger phonon scattering. However, the relationship was inversed for phonons in the frequency range of 1.9 THz to 2.2 THz and higher than 2.5 THz. Phonons in the frequency range 3.0 THz to 3.3 THz had a negative contribution to the heat flux in the P model, but these phonons had a positive contribution in D Models. These high frequency phonons were too localized to conduct heat but their thermal transport capability improved with the introduction of dislocations for two reasons: (1) because of a higher temperature gradient in the D Model, the contribution from high-frequency phonons increased, and (2) the dislocation scattering mostly affected mid-freqeuncy phonons. The dislocations changed the phonon distribution at approximately 2.0 THz such that some phonons with a frequency slightly lower than 2 THz were scattered into phonons with a frequency slightly higher than 2 THz as shown in the phonon density of states (DOS) presented in Supplementary Fig. 2. The spectral heat flux results in the frequency range of 2.2 THz to 2.5 THz did not show a significant dependence on dislocation density. The spectral heat flux $Q(\omega)$ comes from multiplying the phonon distribution $f(\omega)$ by the phonon density of state $D(\omega)$, the phonon group velocity $V_p(\omega)$ and the phonon energy $\hbar\omega$,

$$Q(\omega) = D(\omega)f(\omega)V_p(\omega)\hbar\omega. \tag{3}$$

$D(\omega)$ is an intrinsic material property. $f(\omega)$ was related to the phonon scattering. $V_p(\omega)$ was related to the phonon softening. $Q(\omega)$ was normalized by dividing the total heat flux $Q_{total}$ to better analyze the phonon-dislocation scattering.

The total heat flux $Q_{total}$ can be written as

$$Q_{total} = \int_0^{\omega_{max}} D(\omega)f(\omega)V_p(\omega)\hbar\omega d\omega, \tag{4}$$

The normalized $q(\omega)$ is

$$q(\omega) = \frac{Q(\omega)}{Q_{total}} = \frac{D(\omega)f(\omega)V_p(\omega)\hbar\omega}{\int_0^{\omega_{max}} D(\omega)f(\omega)V_p(\omega)\hbar\omega d\omega} \approx \frac{D(\omega)f(\omega)\hbar\omega}{\int_0^{\omega_{max}} D(\omega)f(\omega)\hbar\omega d\omega}, \tag{5}$$

where $V_p(\omega)$ is weakly frequency independent. The influence from changes in $V_p(\omega)$ was eliminated in $q(\omega)$. The normalized spectral heat fluxes $q(\omega)$ were shown in Fig. 6(b). $Q(\omega)$ and $q(\omega)$ differed the most in the frequency range of 1.1 THz to 1.6 THz where there was change with the dislocation addition in $Q(\omega)$ but no change in the $q(\omega)$ except for a slight decrease when the dislocation density was higher than 6.68 × $10^{15}$ m$^{-2}$. This indicates that the phonon occupancy in the frequency range of 1.1 THz to 1.6 THz did not change, and the $Q(\omega)$ changed in this frequency due to group velocity decrease caused by the strain[42]. The $q(\omega)$ decrease in the highest dislocation density model may have been caused by additional phonon scattering stemming from the interactions between the dislocation and its mirror dislocation across the periodic boundary when the Y scale of the D Model was decreased (see the strain field of the highest dislocation density model as shown in Supplementary Fig. 3). The $q(\omega)$ decreased drastically in the frequency range of 1.6 THz to 2.0 THz and increased in a frequency range of 1.9 THz to 2.2 THz and higher than 2.5 THz in D Model, indicating a strong influence of dislocation scattering. The $q(\omega)$ did not change when the frequency was below 1.6 THz. The phonons in this low frequency range ballistically transported through the model without scattering, since their MFPs were much longer than the length of the simulation.

In summary, by computing the normalized heat flux, we demonstrated that dislocation scattering predominantly affects mid-freqeuncy phonons in the range of 1.6-2.2 THz. The decrease in the spectral heat flux $Q(\omega)$ in the low frequency 1.1-1.6 THz range is likely a result of a group velocity reduction stemming from strain softening. Additionally, the sharp thermal gradient at the dislocation actually facilitates the transport of high frequency phonons with frequencies greater than 2.5 THz. Even though the frequency range of the phonons which were scattered by the dislocation was found, it was still impossible to distinguish the phonon scattering in the LSR and NLSR. So the extended 1D McKelvey-Shockley flux method described in Section 2.2 and 2.3 was used to obtain the phonon MFPs.

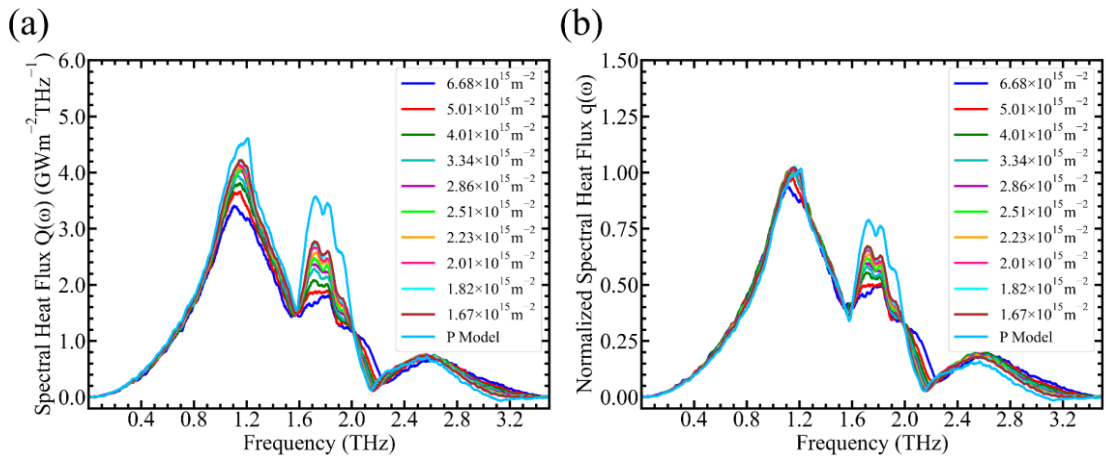

Fig. 6. (a) The spectral heat $Q(\omega)$ of different dislocation density D Model and pristine structure P Model. (b) Normalized spectral heat flux of (a) obtained by dividing the total heat flux.

## 3.3 Spatially resolved phonon dislocation scattering

The phonon mean free paths (MFPs) $\lambda_p(\omega)$ in the P Model and nominal phonon MFPs $\lambda_{nm}(\omega)$ (see Section 2.3) of the three different dislocation density models were obtained. The results are shown in Fig. 7. The length-dependent spectral heat flux results are presented in Supplementary Fig. 4. According to our analysis in the methods Section 2.3, when the phonons experienced extra scattering by the LSR, the $\lambda_{nm}(\omega)$ was larger than the $\lambda_p(\omega)$ in the P Model. When the phonons were scattered equally by the NLSR, the $\lambda_{nm}(\omega)$ was smaller than the $\lambda_p(\omega)$ in the P Model. The results of the medium dislocation density model are shown in Fig. 7(c) and (d). The $\lambda_{nm}(\omega)$ was greater than $\lambda_p(\omega)$ for frequencies greater than 1.9 THz and decreased below $\lambda_p(\omega)$ in the frequency range of 1.6 THz to 1.9 THz. We defined the frequency of the cross-over indicated by yellow dot in Figure 7 as $\lambda_s(\omega)$. Thus, the phonons with frequencies higher than 1.9 THz were scattered by the LSR, while phonon frequencies in the range of 1.6 THz to 1.9 THz were scattered equally or more by the NLSR. The same behavior was observed in the other two dislocation models. However, the corresponding phonon frequency ranges slightly changed. In the highest dislocation density model, $\lambda_s(\omega)$ increased because of the increase in the phonon scattering rate in the LSR, as shown in Fig. 7(a) and (b). Conversely, $\lambda_s(\omega)$ decreased due to the decrease in the phonon scattering rate in the LSR in the lowest dislocation density model, as demonstrated in Fig. 7(e) and (f). The MFPs results were based on the length-dependent heat fluxes and there were very small changes between them in the low-frequency region (< 1.5 THz), as shown in Supplementary Fig. 4. Thus, there were very large fluctuations and big fitting errors in the low-frequency MFPs.

The temperature distribution results demonstrated that in the D Model, the LSR had a much higher phonon scattering rate than the NLSR because of larger temperature gradient in LSR. However, the NLSR accounted for a larger volume than the LSR. Both were important for decreasing the $\kappa_L$. The entire simulation can be interpreted as the dislocation LSR at very high dislocation densities of $10^{16}$ m$^{-2}$ ($1/d^2$, $d$ is the diameter of the LSR). At this condition, we could study the effects on $\kappa_L$ at the upper limit of the dislocation density. Improving the dislocation density higher than $10^{16}$ m$^{-2}$ did not effectively decrease the $\kappa_L$. In this work, the dislocation density was approximately $10^{15}$ m$^{-2}$. Experiments have shown that dislocations have significant impact on the $\kappa_L$ when the dislocation density ranged from $10^{13}$ to $10^{16}$ m$^{-2}$. [20-23, 43] Our simulation results fully matched the experiment.

In summary, our study shows the spatially resolved phonon dislocation scattering processes from the local scattering region and non-local shear scattering region.

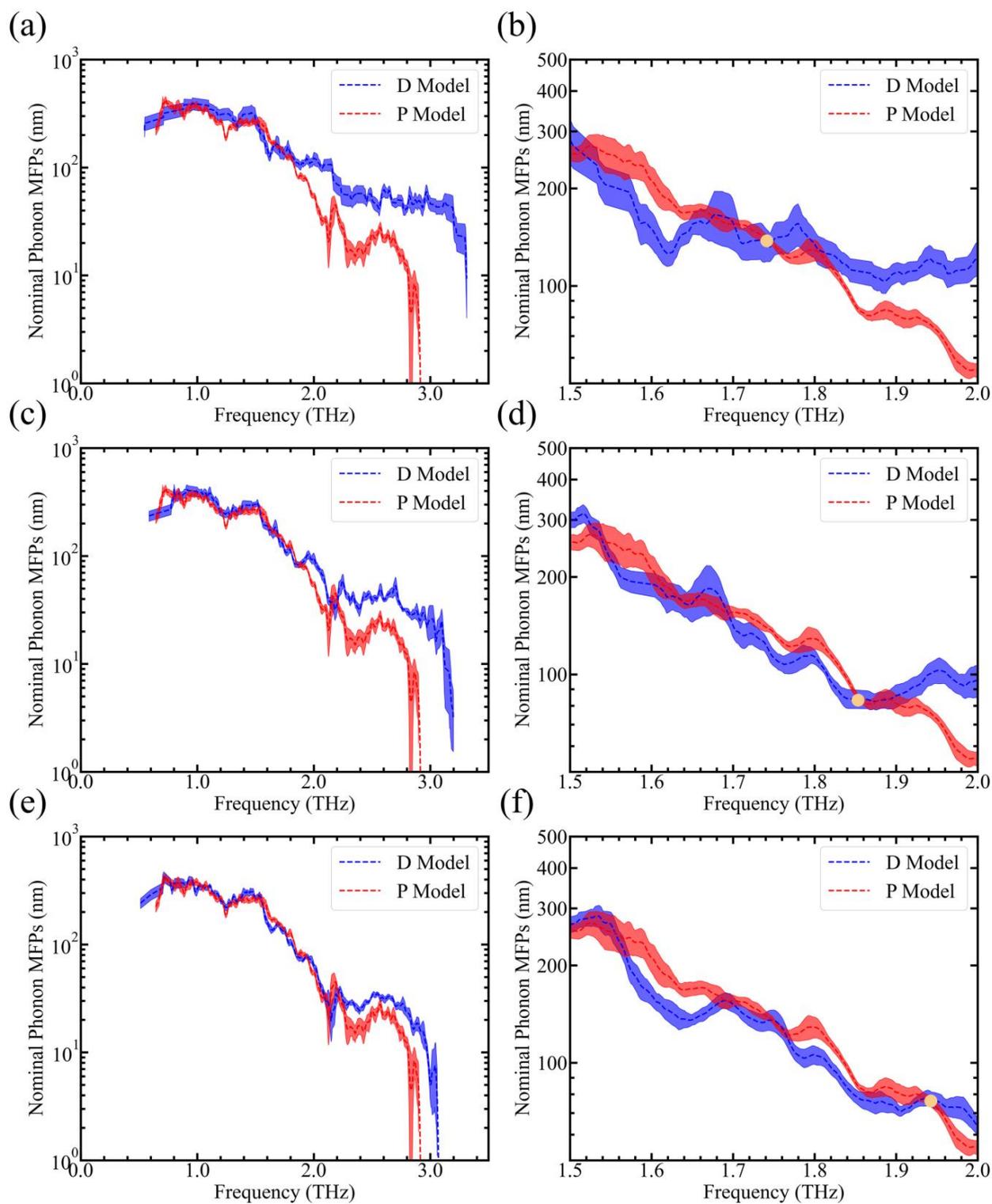

Fig. 7. The nominal phonon MFPs ($\lambda_{nm}(\omega)$) of the dislocation D Model and phonon MFPs ($\lambda_p(\omega)$) of the pristine structure P Model using the 1D McKelvey-Shockley flux method. Of note, the $\lambda_{nm}(\omega)$

is not a real phonon MFPs, a bigger value doesn't mean a weak phonon scattering, but its variations to the phonon MFPs in pristine structure models can reflect the phonon scattering processes in the LSR and NLSR. (a), (c), and (e) are the highest ($6.68 \times 10^{15}$ m$^{-2}$), medium ($3.34 \times 10^{15}$ $m^{-2}$), and lowest ($1.67 \times 10^{15}$ m$^{-2}$) dislocation density model, respectively. (b), (d), and (f) show magnified views of (a) where the $\lambda_{nm}(\omega)$ is smaller than the $\lambda_p(\omega)$.

### 3.4 Shear strain scattering

First-principles calculations were used to verify that the thermal conductivity decreased due to the shear strain in the crystal. Shear strain was added to the pristine PbTe unit cell by changing the orthogonal crystal system into a triclinic crystal system, as shown in Supplementary Fig. 5. This model's shear strain was 2.52% at 20 K as calculated by the MD simulation, which was the same as the shear strain in the medium dislocation density model. The BTE method was applied to calculate the phonon properties using ShengBTE software.

The BTE results are shown in Fig. 8. The cumulative thermal conductivity at 300 K decreased 47.79% from 1.35 W/(m · K) in the pristine model to 0.70 W/(m · K) in the shear strain model (Fig. 8(a)). The contribution from high-frequency phonons was significantly suppressed. However, the group velocity did not obviously change (Fig. 8(b)). The phonon scattering rate is shown in Fig. 8(c). It was higher in the shear strain model than in the pristine model, which was the primary cause of the thermal conductivity reduction in the shear strain model. Shear strain can change the off-diagonal terms of the Grüneisen parameter matrix[44]. Significant increases in the phonons' Grüneisen parameters in frequency ranges of 0.5 THz to 1.6 THz and 2.3 THz to 3.0 THz are shown in Fig. 8(d). Our BTE calculations verified that the crystal's shear strain effectively scattered the phonons and decreased the $\kappa_L$. Thus, the crystal's shear strain stemming from the dislocations contributed to the $\kappa_L$ decrease.

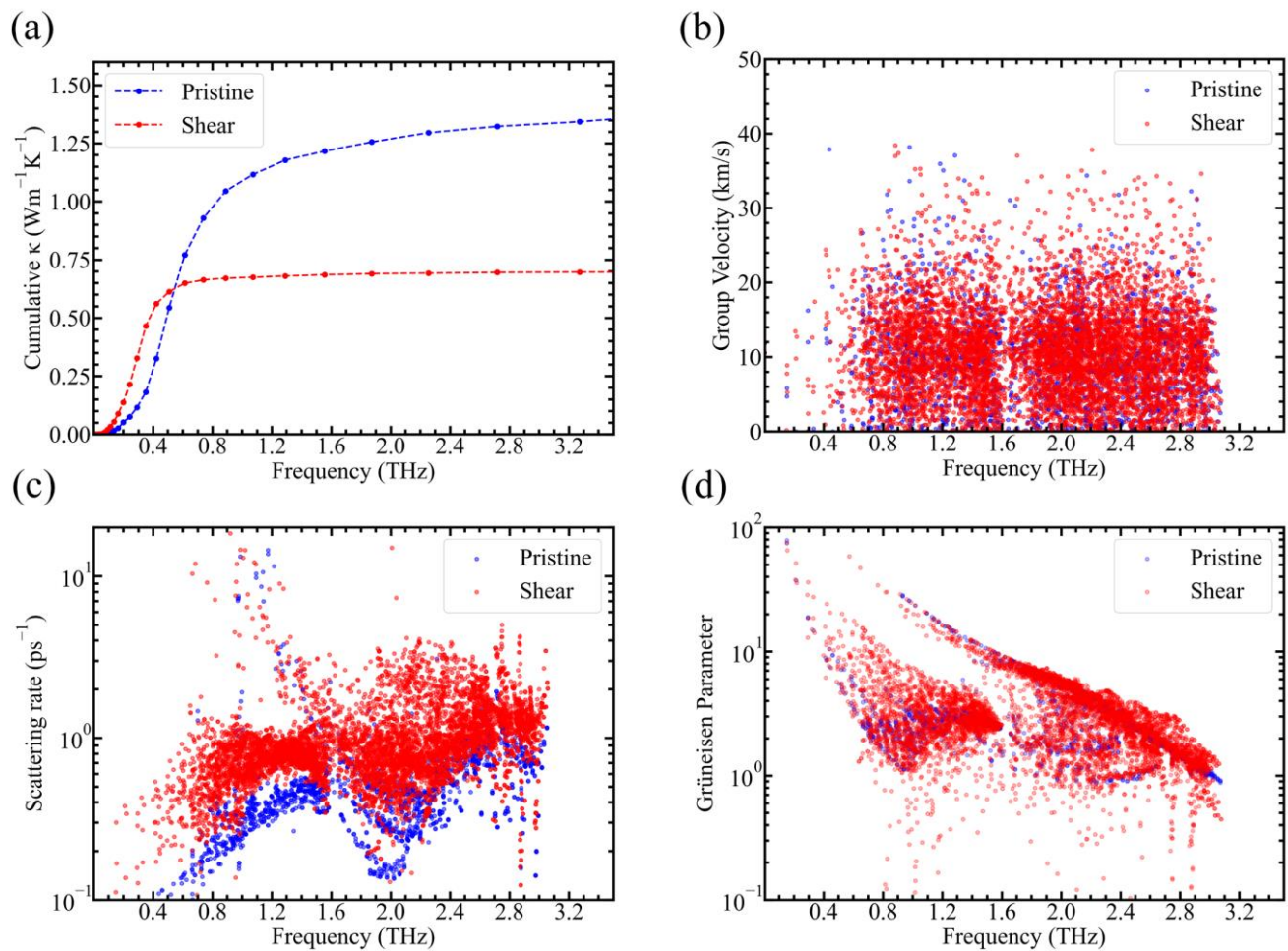

Fig. 8. The phonon properties of the pristine and shear strain model obtained using the BTE calculations: (a) cumulative $\kappa_L$, (b) phonon group velocity, (c) phonon scattering rate, and (d) phonon Grüneisen parameters.

## 4. Conclusion

In this study, the NEMD method was used to assess the lattice thermal conductivity $\kappa_L$ in a PbTe crystal with different dislocation densities. The $\kappa_L$ decreased as the dislocation density increased. Dislocation density ranging from $10^{15}$ to $10^{16}$ m$^{-2}$ was necessary to decrease the $\kappa_L$. Increasing the dislocation density to higher than $10^{16}$ m$^{-2}$ did not effectively decrease the $\kappa_L$. The strain field analysis of the dislocation showed a local scattering region and non-local shear strain scattering region. We separated the phonon scattering from the two regions using the extended McKelvey-Shockley flux method. The non-local shear strain region mainly scattered the medium-frequency phonons and the local scatter region scattered the medium- and high-frequency phonons. The phonon scattering caused by the shear strain was verified by first-principles calculations. Both of the scattering processes were vital for decreasing the $\kappa_L$. To the best of our knowledge, this is the first study demonstrating a spatially resolved phonon dislocation scattering process. This study advances the knowledge of phonon scattering with a complex strain field which will help to regulate the thermal properties of electronic and thermoelectric devices by structure design.

## Supplementary materials

Supplementary note 1

The temperature distributions of the model with dislocation density $6.68 \times 10^{15}$ m$^{-2}$ in addition to the pristine structure P Model in the NEMD simulations, from which we determined the length of the LSR region. It starts from the position where temperature distribution begins to deviate from the linear distribution and ends in the position where temperature distribution comes back to the linear distribution , and the length is the distance in the horizontal direction between the two purple solid circles, see the Fig. 1(b).

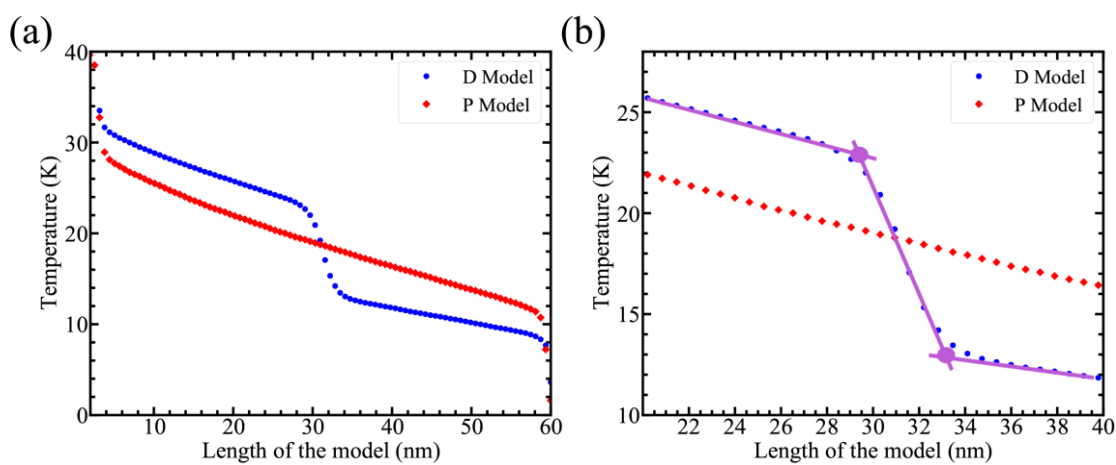

S-Fig. 1. (a) The temperature distributions of the model with dislocation density $6.68 \times 10^{15}$ m$^{-2}$ in addition to the pristine structure P Model in the NEMD simulations. (b) Magnified view showing of the sharp reduction in temperature near the dislocation.

Supplementary note 2

The phonon density of states (DOS) of the dislocation D Model and pristine structure P Model were calculated using the Fourier transform of the velocity autocorrelation function of the atoms in the selected volume containing the dislocation. The position corresponding to this selected volume was used in the P Model. Equilibrium molecular dynamic simulations under 20 K were conducted using the same models in the NEMD simulations.

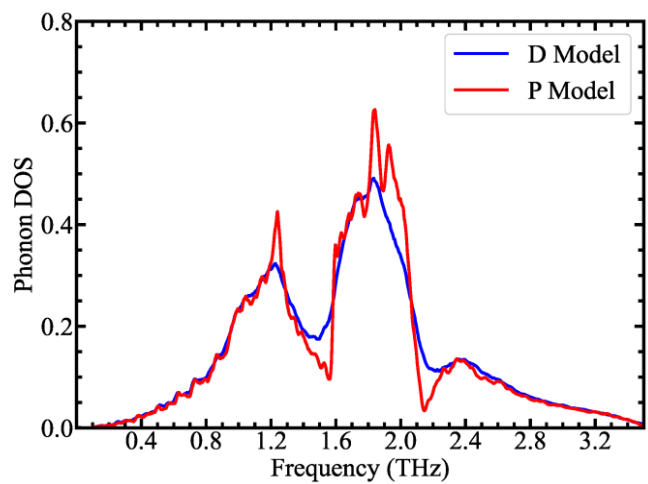

S-Fig. 2. Phonon density of states of the model with dislocation density 3.34 × 10¹⁵ m⁻² D Model and pristine structure P Model.

Supplementary note 3

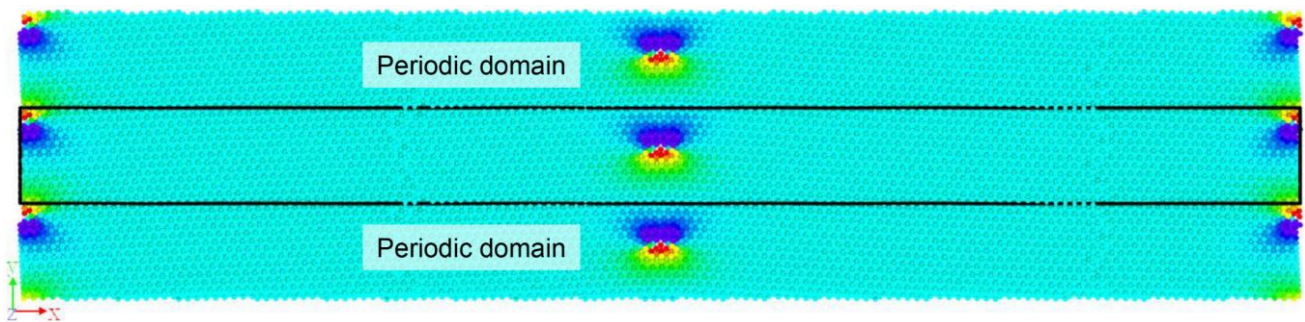

S-Fig. 3. The normal strain $\epsilon_{xx}$ in the the model with dislocation density $6.68 \times 10^{15}$ m$^{-2}$. The negative strain region was close enough to the positive strain region of the nearest periodic domain to allow dislocation interactions.

Supplementary note 4

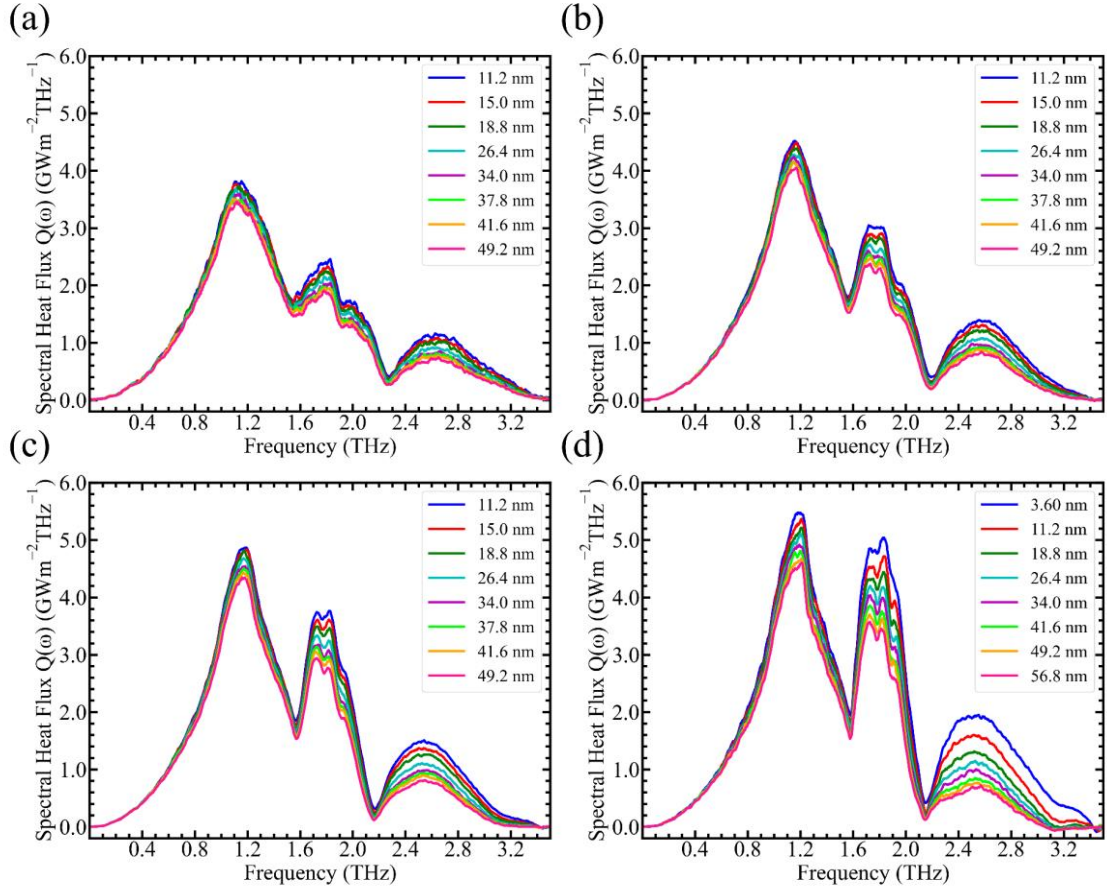

S-Fig. 4. Spectral heat fluxes of the different length models. (a), (b), and (c) Highest (6.68 × $10^{15}$ $m^{-2}$), medium (3.34 × $10^{15}$ $m^{-2}$), and lowest (1.67 × $10^{15}$ $m^{-2}$) dislocation density D Model, respectively. (d) Pristine structure P Model.

Supplementary note 5

A shear strain model was built for first-principles calculations. The Vienna Ab initio Simulation Package (VASP) was used to conduct DFT calculations[45]. The projector augmented wave (PAW)[46] method was used in conjunction with the revised Perdew-Burke-Ernzerhof for solids (PBEsol)[47] generalized gradient approximation (GGA)[48] for the exchange-correlation (xc) functional[49]. For structural relaxation, 16 × 16 × 16 Monkhorst-Pack k-point meshes and a plane wave basis with a kinetic energy cut-off of 330 eV were used. The force and energy convergence thresholds were $10^{-8}$ eV/Å. The configurations for calculating the second- and third-order force constants were generated by Phonopy and thirdorder.py in the ShengBTE package, respectively. A 2 × 2 × 2 unit cell was used. The forces on the atoms were calculated using VASP, and the second- and third-order force constants were obtained by Phonopy and thirdorder.py in the ShengBTE package, respectively. The phonon dispersion was obtained from the second- and third-order force constants. By combining the second- and third-order force constants, the BTE was solved using the ShengBTE package to obtain the thermal conductivity, phonon group velocity, phonon scattering rate, and Grüneisen constant.

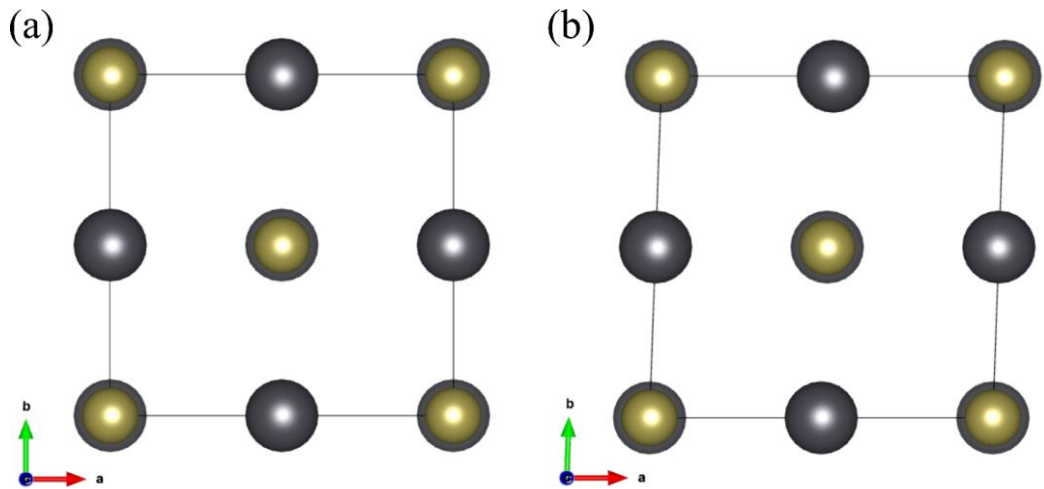

S-Fig. 5. Shear strain model built using first-principles calculations: (a) pristine model and (b) shear strain model. The angle between the **a** and **b** axis was 88°. The figure was produced using VESTA software[50].

## Data availability

The datasets generated during and/or analyzed during this study are available from the corresponding author on reasonable request.

## Code availability

The codes used in this study are available from the corresponding author on reasonable request.

## Acknowledgment

Y.D.S. thanks Dr. Shiyao Liu for polishing the language. Simulations were performed with computing resources granted by the National Supercomputer Center in Tianjin under project TianHe-1(A) and Quest High-Performance Computing Cluster at Northwestern University. Ben Xu acknowledges support from the NSFC under grant No. 52072209, Basic Science Center Project of NSFC under grant No.51788104 and NSAF No. U1930403. G. Jeff Snyder acknowledges support from award 70NANB19H005 from U.S. Department of Commerce, National Institute of Standards and Technology as part of the Center for Hierarchical Materials Design (CHiMaD).

## Author contributions

Y.D.S. performed MD simulations, data processing, result analysis, and manuscript writing. J.Y.Z contributed to building the dislocation net model. Y.G.Z., R. M., M.H., B.X., W.L., and J. S. participated in the discussion and interpretation of results and contributed to editing the manuscript.

Competing interests

The authors declare no competing interests.